\documentclass[aps,prd,twocolumn,floatfix,preprintnumbers,notitlepage,nofootinbib,longbibliography]{revtex4-1}
\usepackage[dvipdfmx]{graphicx}
\usepackage{color}
\usepackage{natbib}
\usepackage{amsmath}
\usepackage{amsfonts}
\usepackage[colorlinks=true,citecolor=blue,urlcolor=blue]{hyperref}

\newcommand{\km}{\mathrm{km}}
\newcommand{\s}{\mathrm{s}}
\newcommand{\Mpc}{\mathrm{Mpc}}
\newcommand{\eV}{\mathrm{eV}}

\begin{document}
\title{The implications of an extended dark energy cosmology with massive neutrinos for cosmological tensions}
\author{Vivian Poulin, Kimberly K. Boddy, Simeon Bird, and Marc Kamionkowski}
\affiliation{Department of Physics and Astronomy, Johns Hopkins University, Baltimore, MD 21218, USA}
\affiliation{Department of Physics and Astronomy, UC Riverside, Riverside, CA 92512, USA}

\begin{abstract}


  We perform a comprehensive analysis of the most common early- and late-Universe solutions to the $H_0$, Ly-$\alpha$, and $S_8$ discrepancies.
  When considered on their own, massive neutrinos provide a natural solution to the $S_8$ discrepancy at the expense of \textit{increasing} the $H_0$ tension.
  If all extensions are considered simultaneously, the best-fit solution has a neutrino mass sum of $\sim 0.4$~eV, a dark energy equation of state close to that of a cosmological constant, and no additional relativistic degrees of freedom.
  However, the $H_0$ tension, while weakened, remains unresolved.
  Motivated by this result, we perform a non-parametric reconstruction of the evolution of the dark energy fluid density (allowing for negative energy densities), together with massive neutrinos.
  When all datasets are included, there exists a residual $\sim1.9\sigma$ tension with $H_0$.
  If this residual tension remains in the future, it will indicate that it is not possible to solve the $H_0$ tension solely with a modification of the late-Universe dynamics within standard general relativity.
  However, we do find that it is possible to resolve the tension if \textit{either} galaxy BAO \textit{or} JLA supernovae data are omitted.
  We find that \textit{negative} dark energy densities are favored near redshift $z\sim2.35$ when including the Ly-$\alpha$ BAO measurement (at $\sim 2\sigma$).
  This behavior may point to a negative curvature, but it is most likely indicative of systematics or at least an underestimated covariance matrix.
  Quite remarkably, we find that in the extended cosmologies considered in this work, the neutrino mass sum is always close to $0.4~\eV$ regardless of the choice of external datasets, as long as the $H_0$ tension is solved or significantly decreased.

\end{abstract}
\maketitle

\section{Introduction}

The concordance $\Lambda$CDM model of cosmology is very successful in explaining the large-scale structure (LSS) of the Universe; it passes a number of precision tests and describes well observations of the cosmic microwave background (CMB) from the \textit{Planck} satellite~\cite{Ade:2015xua}.
However, with the increasing precision and sensitivity of various instruments, interesting tensions have emerged.
A recent direct measurement of the local value of the present day Hubble rate $H_0$~\cite{Riess:2016jrr} shows a $> 3\sigma$ tension with the inferred value from CMB observations~\cite{Ade:2015xua}.
Furthermore, there is a long-standing discrepancy between LSS surveys and the CMB determination of the quantity $S_8=\sigma_8(\Omega_M/\Omega_M^\mathrm{ref})^{\alpha}$, where $\sigma_8$ is the amplitude of matter density fluctuations in spheres with radius of $8h^{-1}~\Mpc$, $\Omega_M$ is the relic density of matter in the Universe today, and $\Omega_M^\mathrm{ref}$ is a normalization.%
\footnote{The values of $\Omega_M^\mathrm{ref}$ and $\alpha$ vary between experiments, but they are often set to $0.3$ and $0.5$, respectively.}
Measurements of $S_8$ from galaxy clustering and weak lensing surveys (such as CFHTLenS~\cite{Heymans:2013fya}, KiDS~\cite{Hildebrandt:2016iqg,Kohlinger:2017sxk}, DES~\cite{Abbott:2017wau}, and Planck SZ cluster counts~\cite{Ade:2015fva}) are all smaller (between 2$\sigma$ and 4$\sigma$) than the CMB prediction.
Finally, the BOSS DR11 baryon acoustic oscillation (BAO) measurements from the Ly-$\alpha$ auto-correlation analysis and cross-correlation with quasars have a reported $\sim 2.5\sigma$ tension with the flat $\Lambda$CDM \textit{Planck} prediction~\cite{Delubac:2014aqe}.
The significance of this discrepancy is reduced by recent increases in the size of the dataset, perhaps suggesting a statistical fluctuation combined with a mildly non-Gaussian covariance matrix~\cite{Bautista:2017zgn}, but a 2.3$\sigma$ tension remains with the latest DR12 data~\cite{Bourboux:2017cbm}.
There have been various efforts to resolve these tensions with different cosmological models, usually classified as either early- or late-Universe solutions~\cite{Berezhiani:2015yta,Chudaykin:2016yfk,Poulin:2016nat,DiValentino:2017oaw,DiValentino:2016hlg,DiValentino:2017zyq,DiValentino:2017iww,DiValentino:2017rcr,Addison:2017fdm,Buen-Abad:2017gxg,Raveri:2017jto,Mortsell:2018mfj}.
These attempts often focus on solving one of the tensions, using specific datasets to fit simple extensions of $\Lambda$CDM.
However, these extensions are inconsistent when additional datasets constraining late-Universe expansion quantities, such as the BAO scale or the luminosity distance from type Ia supernovae (SNe Ia), are incorporated~\cite{Riess:2016jrr,DiValentino:2017zyq,Addison:2017fdm}.

In this paper, we consider a wide range of datasets measuring both the early- and late-Universe properties to see if a coherent model emerges.
We focus on massive neutrino solutions\footnote{Another class of potential solutions involves interacting~\cite{DiValentino:2017oaw,Buen-Abad:2017gxg,Raveri:2017jto} or decaying dark matter~\cite{Berezhiani:2015yta,Chudaykin:2016yfk,Poulin:2016nat} in an isolated dark sector.} to the $S_8$ problem, because they are the less ``theoretically costly'': oscillation experiments indicate that neutrinos must have non-zero masses.
Moreover, massive neutrinos reduce the growth of perturbations below their free-streaming length~\cite{Lesgourgues:2006nd}, and dedicated studies point to a neutrino mass sum $\sum m_\nu\sim0.4~\eV$~\cite{Wyman:2013lza,Battye:2013xqa,Beutler:2014yhv,McCarthy:2016mry,McCarthy:2017csu}.
Unfortunately, such a solution is in apparent conflict with the local $H_0$ measurements: the value of $\sum m_\nu$ results in a \textit{lower} Hubble rate inferred from the CMB, ultimately exacerbating the $H_0$ tension.
We approach the $H_0$, Ly-$\alpha$, and $S_8$ tensions in two ways.
We first attempt to solve all tensions simultaneously by combining the most common early- and late-Universe extensions of $\Lambda$CDM.
We incorporate massive neutrinos, and we allow for an additional ultra-relativistic species with $\Delta N_\mathrm{eff}$ and an arbitrary effective sound speed $c_\mathrm{eff}^2$ and viscosity speed $c_\mathrm{vis}^2$.
We model the dark energy (DE) sector as a fluid whose equation of state is given by the CPL parameterization $w(a)=w_0 + (1-a)w_a$~\cite{Chevallier:2000qy}.
Using \textit{Planck} CMB data~\cite{Aghanim:2015xee}, \textit{Planck} SZ data~\cite{Ade:2015fva}, and the recent $H_0$ measurement~\cite{Riess:2016jrr}, we find that resolving the $H_0$ and $S_8$ tensions \textit{simultaneously} require phantom-like DE~\cite{Caldwell:1999ew} and $\sum m_\nu \sim 0.4~\eV$.
However, this conclusion is spoiled when external galaxy BAO or SNe Ia data are included, even in the presence of an additional relativistic fluid.

Given this persistent inconsistency, we perform an agnostic reconstruction of an exotic DE sector (ExDE) to determine the dynamics necessary to reconcile problematic low-redshift data with other cosmological probes.
While there have been similar approaches with phenomenological reconstructions of the Hubble parameter $H(z)$~\cite{Bernal:2016gxb} and the DE equation of state $w(z)$~\cite{Zhao:2017cud}, our analysis differs in several ways.
In the former analysis~\cite{Bernal:2016gxb}, only data measuring the late-Universe expansion are considered.
This requires a prior on the sound horizon at baryon drag $r_s^\textrm{drag}$ and diminishes the constraining power on the matter and baryon energy densities, $\omega_m$ and $\omega_b$.
In the latter analysis~\cite{Zhao:2017cud}, the behavior of $w(z)$ strongly deviates from the nominal case of a cosmological constant with $w=-1$ in a manner that is not captured by the CPL parameterization.
However, by only modifying the equation of state, the energy density of the fluid is necessarily positive.
In our reconstruction, we allow the energy density $\Omega_\mathrm{ExDE}(z)$ to take on both positive and negative values.
Although we assign this energy density to the DE sector, it can also be thought of as a proxy for any number of new species that could collectively give rise to the arbitrarily complicated dynamics favored by the CMB and low-redshift data.
Hence, it can indicate that the energy density in another sector must decrease (as is the case, for instance, if part of the dark matter is decaying or if the Universe has an open geometry).
Naturally, this can also indicate a strong inconsistency in the data.

With our formalism, we are able to solve the $H_0$, Ly-$\alpha$, and $S_8$ tensions and achieve compatibility with the CMB, LSS, and \textit{either} galaxy measurements of the BAO scale \textit{or} measurements of SNe Ia.
There is a $\sim1.9\sigma$ tension with $H_0$ that persists when all datasets are included in our analysis, a finding consistent with previous studies~\cite{Bernal:2016gxb,Zhao:2017cud}.
This is because the BAO and SNe Ia data prefer slightly different expansion histories at late times, ultimately forcing the behavior of the ExDE to be very close to that of a cosmological constant below $z<0.6$.
If this residual tension remains in the future, it would indicate that it is not possible to solve the $H_0$ tension solely with a modification of the late-Universe dynamics within standard general relativity.
We have additionally allowed for an extra ultra-relativistic fluid, but it neither affects the reconstruction nor helps reduce the tension.
Moreover, we find that the Ly-$\alpha$ BAO measurements favor \textit{negative} values of $\Omega_\mathrm{ExDE}(z)$ at $z \sim 2.5$.
We discuss possible explanations of such behavior, but stress that this may point to systematics in the data.
Last but not least, we find that the neutrino mass sum is close to $0.4~\eV$, regardless of the choice of external datasets, as long as the $H_0$ tension is solved or significantly decreased.
We have verified that this finding remains true when including $A_\textrm{lens}$ as a free parameter~\cite{Calabrese:2008rt}.

This paper is organized as follows.
Section~\ref{sec:prelim} is devoted to a preliminary discussion on the $H_0$ and $S_8$ tensions and particular solutions.
We perform an in-depth analysis of a combination of the most common extensions to $\Lambda$CDM advocated to solve these tensions in Section~\ref{sec:combine}, followed by an agnostic approach in Section~\ref{sec:agnostic}.
From this reconstruction, we discuss in Section~\ref{sec:discussion} models that explain this behavior and therefore provide a solution to the $S_8$, $H_0$, and Ly$-\alpha$ tensions without spoiling the successful description of other probes.

\section{Preliminary Considerations}
\label{sec:prelim}

In this section, we discuss how the present-day Hubble rate $H_0$ and the quantity $S_8$ are measured or inferred from observations, and we comment on the discrepancies seen between experiments.
We then discuss the standard extentions of $\Lambda$CDM that are most often invoked in the attempt to reconcile these discrepancies.
Although certain cosmological models may lessen tensions with specific data sets, no solutions are robust to the inclusion of additional datasets such as the BAO or SNe Ia.

\subsection{Datasets and analysis procedure}
\label{sec:datasets}

We summarize the various datasets considered in the remainder of this work.
\begin{itemize}
\item CMB: In Section~\ref{sec:combine}, we use the \textit{Planck} 2015 high-$\ell$ TT, TE, and EE power spectra~\cite{Aghanim:2015xee} with a gaussian prior on $\tau_\mathrm{reio} = 0.055\pm 0.009$, given by the SIMlow likelihood~\cite{Aghanim:2016yuo}.
  We also include the \textit{Planck} lensing likelihood~\cite{Ade:2015zua}.
  In Section~\ref{sec:agnostic}, we instead use the lite version of this dataset to decrease the convergence time of our likelihood analysis.
  We have verified that doing so has no impact on our conclusions, apart from slightly increasing the error bars on the fitted cosmological parameters.
\item LSS: We use the measurement of the halo power spectrum from the Luminous Red Galaxies SDSS-DR7~\cite{Reid:2009xm} and the full correlation functions from the CFHTLenS weak lensing survey~\cite{Heymans:2013fya}.
  We also use the $S_8$ measurement from the \textit{Planck} SZ cluster counts~\cite{Ade:2013lmv}, since it is at the heart of the claimed $S_8$ discrepancy.
  Although not included in our likelihood analysis, we later assess whether our best fit model can accommodate the $S_8$ measurements from KiDS~\cite{Kohlinger:2017sxk} and DES1~\cite{Troxel:2017xyo}.
\item SH0ES: We use the SH0ES measurement of the present-day Hubble rate $H_0=73.24\pm0.174$~\cite{Riess:2016jrr}.
\item BAO: We use measurements of the volume distance from 6dFGS at $z = 0.106$~\cite{Beutler:2011hx} and the MGS galaxy sample of SDSS at $z = 0.15$~\cite{Ross:2014qpa}, as well as the recent DES1 BAO measurement at $z=0.81$~\cite{Abbott:2017wcz}.
  We include the anisotropic measurements from the CMASS and LOWZ galaxy samples from the BOSS DR12 at $z = 0.38$, $0.51$, and $0.61$~\cite{Alam:2016hwk}.
  The BOSS DR12 measurements also include measurements of the growth function $f$, defined by
  \begin{equation}
    f\sigma_8 \equiv \frac{\left[\sigma_8^{(vd)}(z)\right]^2}
                          {\sigma_8^{(dd)}(z)} \ ,
  \end{equation}
  where $\sigma_8^{(vd)}$ measures the smoothed density-velocity correlation, analogous to $\sigma_8 \equiv \sigma_8^{(dd)}$ that measures the smoothed density-density correlation.
\item Ly-$\alpha$: The latest lyman-$\alpha$ BAO (auto and cross-correlation with quasars) at $z=1.5$~\cite{Ata:2017dya}, $z=2.33$~\cite{Bautista:2017zgn} and $z=2.4$~\cite{Bourboux:2017cbm} are not yet public, but are known to be in slightly better agreement with $\Lambda$CDM than the DR11 data.
  We therefore incorporate them in the form a Gaussian likelihood and have verified that it gives similar results as the full DR11 likelihood~\cite{Font-Ribera:2013wce,Delubac:2014aqe}.
\item JLA: We use the SDSS-II/SNLS3 Joint Light-Curve Analysis (JLA) data compilation of $> 740$ SNe Ia at redshifts $0.01 \lesssim z \lesssim 1.3$~\cite{Betoule:2014frx}.
\end{itemize}

Our primary analysis includes all datasets simultaneously, since our goal is to try to find a coherent cosmological model that can explain seemingly incompatible data.
Using the public code \texttt{Monte Python}~\cite{Audren12}, we run Monte Carlo Markov chain analyses with the Metropolis-Hastings algorithm and assume flat priors on all parameters.
Our $\Lambda$CDM parameters are
\begin{equation*}
  \{\omega_\mathrm{cdm},\omega_b,\theta_s,A_s,n_s,\tau_\mathrm{reio}\} \ .
\end{equation*}
There are many nuisance parameters for the \textit{Planck}~\cite{Aghanim:2015xee} and JLA~\cite{Betoule:2014frx} likelihoods that we analyze together with these cosmological parameters.\footnote{For the nuisance parameters, we use the default priors that are provided by \texttt{MontePython}.}
We use a Cholesky decomposition to handle the large number of nuisance parameters~\cite{Lewis:2013hha}.
Using the Gelman-Rubin criterion~\cite{Gelman:1992zz}, we apply the condition $R -1<0.05$ to indicate our chains have converged.

\subsection[The discrepancy between local H0 and the CMB]{The discrepancy between local distance measurements of $H_0$ and the CMB}
\label{sec:prelim_H0}

Observations of the CMB provide a firm measurement of the distance scale at decoupling:
\begin{align}
  d_s(z_\mathrm{dec}) &= \frac{1}{1+z_\mathrm{dec}}
  \int_{z_\mathrm{dec}}^{\infty} \frac{c_s}{H(z)} \; dz \ .
  \label{eq:d_s}
\end{align}
This represents an early-time anchor of the cosmic distance ladder.
The CMB also provides an estimate of a late-time anchor of the distance ladder: $H_0$, the expansion rate today (see, \textit{e.g.}, Chapter~5.1 in Ref.~\cite{lesgourgues2013neutrino} for more details).
However, this measurement is indirect and depends on the assumed cosmological model.
Thus, the \textit{direct} determination of $H_0$ at low-redshift is essential to firmly calibrate the distance ladder in a model independent fashion.

The SH0ES survey measured the value of the present-day Hubble rate to a precision of $2.4\%$,
by constructing a local cosmic distance ladder from Cepheids and supernovae at $z<0.15$.
Their final result is $H_0 = 73.24\pm1.74~\km/\s/\Mpc$~\cite{Riess:2016jrr}.
This direct measurement of $H_0$ is discrepant at the $\sim3.4\sigma$ level with the inferred value of $H_0 = 66.93 \pm 0.62~\km/\s/\Mpc$ from \textit{Planck}~\cite{Aghanim:2016yuo} (from the TT+TE+EE+SIMlow measurements at the 68\% confidence level).


\subsubsection{Early-time solutions}

To resolve the tension between the \textit{Planck} and SH0ES determination of $H_0$ within $\Lambda$CDM by modifying the distance ladder at early times, the CMB-inferred value of $d_s(z_\mathrm{dec})$ must be reduced by a factor of $\sim 6\%$ to $10~\Mpc$~\cite{Bernal:2016gxb}.
As a result, either the sound speed in the photon--baryon plasma must decrease or the redshift of recombination must increase [see Eq.~\eqref{eq:d_s}].
To achieve these effects, a higher primordial helium fraction $Y_p$ or an extra ultra-relativistic species are often invoked.%
\footnote{In principle, any species affecting the background expansion at early times could be used. See, \textit{e.g.}, Ref.~\cite{Karwal:2016vyq} for an alternative attempts at solving the $H_0$ discrepancy via an early DE component.}
However, both these possibilities are ruled out.
The CMB and big-bang nucleosynthesis (BBN) constrain $Y_p$ to be close to $0.25$~\cite{Bernal:2016gxb}.
Extra relativistic degrees of freedom sufficient to recover the low-redshift value of $H_0$ are ruled out within $\Lambda$CDM by \textit{Planck} polarization data and BAO measurements~\cite{Bernal:2016gxb,DiValentino:2017iww}.

\subsubsection{Late-time solutions}

Late-time solutions for this discrepancy rely on altering the expansion history, such that the expansion rate matches the CMB at decoupling and the local rate today.

Within $\Lambda$CDM it is not possible to accomodate both $H_0$ and BAO data, which fix the expansion history between $z=2.3$ and $z=0.15$ ; the only extra low-redshift degree of freedom is the ratio between $\Omega_\Lambda$ and $\Omega_m$, which is insufficient to allow the expansion history to change significantly between $z=0.15$ and $z=0$.

Alternative standard extensions attempting to solve the $H_0$ discrepancy include a phantom-like dark energy (DE) component with an equation of state $w<-1$~\cite{DiValentino:2016hlg,DiValentino:2017zyq}, a vacuum phase transition~\cite{DiValentino:2017rcr}, or interacting DE~\cite{DiValentino:2017iww}.
However, assuming an early time cosmology as in $\Lambda$CDM, it is hard to reconcile these possible solutions with BAO data and JLA data \cite{DiValentino:2016hlg,DiValentino:2017zyq,DiValentino:2017rcr,DiValentino:2017iww}.

In conclusion, when considered separately from each other, the most common extensions to the standard cosmological model are too tightly constrained to explain the tension with local $H_0$ measurements if BAO and SNe Ia data in agreement with \textit{Planck} are included in the analysis.

\subsection{The discrepancy between the power spectrum amplitude from the CMB and LSS}
\label{sec:prelim_S8}

There is a moderate tension within $\Lambda$CDM between the value of $S_8$ measured by LSS survey
Galaxy clustering and weak lensing surveys (such as CFHTLenS~\cite{Heymans:2013fya}, KiDS~\cite{Hildebrandt:2016iqg,Kohlinger:2017sxk}, DES~\cite{Abbott:2017wau}, and \textit{Planck} SZ cluster counts~\cite{Ade:2015fva}) measure a value of $S_8$ between $2\sigma$ and $4\sigma$ smaller than that inferred from the CMB.
Note that, through lensing, the CMB measures the power spectrum amplitude not only at $z=1100$, but also over a redshift range centred at $z \approx 2$.
These two \textit{Planck} measurements are internally inconsistent, and the nuisance parameter $A_\textrm{lens}$ is used to allow them to vary freely.
Marginalising over $A_\textrm{lens}$ reduces the significance of the $S_8$ tension but does not remove it, because the lensing 4-point correlation estimator $C_l^{\rm \phi\phi}$ itself does not favor high value of $A_\textrm{lens}$.
Indeed, the amount of lensing measured from the smoothing of high multipole peaks in the TT spectrum is higher than that measured from $C_l^{\rm \phi\phi}$, the latter being compatible with the $\Lambda$CDM expectation~\cite{Ade:2015fva,Ade:2015zua}.
Weak lensing measurements probe a lower redshift range, $z \approx 0.4 - 1.0$, compared to CMB lensing.
Furthermore, weak lensing surveys and galaxy clusters measure $S_8$ on smaller scales than the \textit{Planck} CMB, $k\sim0.1$ Mpc and $\sim 8$ Mpc, respectively.

This motivates solutions that change either the growth rate of structure for $z < 2$ or alter the shape of the power spectrum on small scales \cite{Kamionkowski:1999vp, Sigurdson:2003vy}.
Interactions in the dark matter sector helps to address the $S_8$ problem~\cite{Lesgourgues:2015wza,Buen-Abad:2017gxg,Raveri:2017jto}, but are in tension with Ly-$\alpha$ data~\cite{Buen-Abad:2017gxg,Krall:2017xcw}.
Here, we focus on another possibility; massive neutrinos, which reduce power on small scales by reducing the growth rate.

\subsubsection{Solutions due to massive neutrinos}

There is some weak evidence from cosmology supporting a non-zero neutrino mass sum.
For example, Ref.~\cite{Beutler:2014yhv} found a $2.6\sigma$ preference for a non-zero neutrino mass from SDSS, and $S_8$ constraints from galaxy cluster counts give similar results~\cite{Wyman:2013lza,Battye:2013xqa}.
Recently, Ref.~\cite{McCarthy:2017csu} combined \textit{Planck} CMB measurements with thermal Sunyaev-Zeldovich (tSZ), BAO, and lensing data.
They used a suite of hydrodynamic simulations calibrated to produce realistic cluster gas profiles~\cite{McCarthy:2016mry}.
Central to their analysis was removing the internal tension between \textit{Planck} CMB and \textit{Planck} lensing by marginalising over $A_\mathrm{lens}$.
Their conclusions are in striking agreement with those of this work, finding that a neutrino mass sum $\sum m_\nu \sim 0.4~\eV$ is preferred by most tSZ and lensing effects, although details of their analysis made a formal significance challenging.
Although we do not directly include tSZ data here, we note that it would only strengthen our conclusions about neutrino masses.

There are also some datasets which appear to rule out a neutrino mass sum of the value preferred by our analysis. 
Most notably, the small-scale 1D Ly-$\alpha$ forest flux power spectrum can be combined with \textit{Planck} to constrain the neutrino mass sum to be $\sum m_\nu < 0.12~\eV$~\cite{PalanqueDelabrouille:2015pga}.
Note that the forest alone constrains only $\sum m_\nu < 1~\eV$.
As the Ly-$\alpha$ forest is sensitive to the matter power spectrum on non-linear scales of $k = 0.1$--$4\, h/\Mpc$, this constraint requires simulations for calibration and assumes a $\Lambda$CDM cosmology.
Given that our models include substantial deviations from $\Lambda$CDM even at $z > 2$, along with the lack of a public likelihood function code, we chose not to use this Ly-$\alpha$ forest dataset.

However, we note that the Ly-$\alpha$ forest measures a spectral index $n_s = 0.9238 \pm 0.01$, $2$--$3\sigma$ lower than the $n_s = 0.9655 \pm 0.0062$ from \textit{Planck}~\cite{PalanqueDelabrouille:2015pga,Ade:2015xua}.
Thus, the Ly-$\alpha$ forest, in agreement with the rest of our analysis, does prefer reduced power on small scales compared to the CMB.
A Ly-$\alpha$ forest analysis allowing for a more general dark energy model would be an interesting check on our conclusions, and we may address this in future work.
We also note that constraints on $\sum m_\nu$ usually depends on the assumed DE equation of state; they can be very strong when $w\geq1$ is assumed (see e.g. the recent \cite{Giusarma:2018jei,Vagnozzi:2018jhn}), but largely relaxes when negative $w$ (as favored by the combination of CMB and SHOES data) are allowed \cite{Vagnozzi:2018jhn}.

\section[Combining the most common extensions to LCDM]{Combining the most common extensions to $\Lambda$CDM}
\label{sec:combine}

We have argued that the most common extensions to $\Lambda$CDM invoked in order to solve the $H_0$ and $S_8$ problems, when considered separately, are not able to accommodate all datasets currently available.
In this section, we consider a combination of these extensions to see if they can achieve in concert what they could not alone.
We retain the basic framework of $\Lambda$CDM throughout this section, considering only well-motivated extensions.

\subsection{Models}

We denote the standard $\Lambda$CDM cosmology with massless neutrinos as $\nu_0\Lambda$CDM, and we consider the following modifications:
\begin{itemize}
\item Massive neutrinos: We consider a degenerate mass hierarchy for the neutrinos, as we find the specification of the mass hierarchy to be irrelevant for current datasets.
  The exception is if one of the neutrinos is massless, in which case the matter power spectrum is significantly altered~\cite{Lesgourgues:2006nd}.
  
\item DE as a scalar field: We use the CPL parameterization $w(a)=w_0+(1-a)w_a$~\cite{Chevallier:2000qy}, with a parameterized post-Friedmann treatment to allow the crossing of the phantom divide~\cite{Fang:2008sn}.
  We set the sound speed in the rest frame of the scalar field to unity and use the priors $w_0\in[-3,0.3]$ and $w_a\in [-2,2]$~\cite{DiValentino:2017zyq}.
\item Additional ultra-relativistic species: There are many models that introduce additional relativistic degrees of freedom $\Delta N_\mathrm{eff}$.
  For example, extra active or sterile neutrinos, light scalar fields, or dark radiation in a dark sector.
  For a given $\Delta N_\mathrm{eff}$, all of these models have the same \textit{background} effects on the CMB, but there are a number of \textit{perturbation} effects that are model dependent (for instance, a free-streaming species is known to induce a shift of shifts CMB peaks towards larger scales, or smaller angles---an effect known as "neutrino drag").

  To keep the discussion as general and model-independent as possible, there is a postulated linear and time-independent relation between the isotropic pressure perturbations and density perturbations $\delta p/\delta\rho= c^2_\mathrm{eff}$ (defined in the rest frame of the ultra-relativistic species); similarly, there is a viscosity coefficient $c^2_\mathrm{vis}$ that enters the source term of the anisotropic pressure~\cite{Hu:1995em,Audren:2014lsa,Ade:2015fva,Bernal:2016gxb}.
  We add an ultra-relativistic species, which does not share the same mass as the active neutrinos, by modifying $N_\mathrm{eff}$, the effective sound speed $c^2_\mathrm{eff}$, and the viscosity sound speed $c^2_\mathrm{vis}$.
  We use the priors $\Delta N_\mathrm{eff} \in [-1,1]$ and $c^2_\mathrm{eff}, c^2_\mathrm{vis} \in [0,1]$.
\end{itemize}
We refer to the model combining all these extensions as $\nu_Mw$CDM+$N_\mathrm{fluid}$.

\subsection{Results}

\subsubsection{Restricted Datasets}

First, we perform an analysis that includes only the CMB, the SH0ES, and \textit{Planck} SZ datasets.
With these datasets alone, an extended model can solve the tension between the CMB and SH0ES and the tension between the CMB and \textit{Planck} SZ \textit{simultaneously}.
We find $H_0 = 72.6\pm1.8$, in agreement with local measurements, while $(\sigma_8,\Omega_M)=(0.7823_{-0.017}^{+0.017},0.2862_{-0.016}^{+0.014})$, in agreement with the low-$z$ measurements.
This is possible because the extra freedom allowed by our extended cosmological model is absorbed by the CMB.
What was previously a tension thus appears as extended parameters which deviate strongly from $\Lambda$CDM.
We have a neutrino mass sum $\sum m_\nu  = 0.67_{-0.17}^{+0.13}~\eV$ and DE parameters $(w_0,w_a)=(-1.205_{-0.23}^{+0.13},-1.492_{-1.00}^{+0.34})$.
The goodness of fit is $\Delta \chi^2_\mathrm{min}=\chi^2_\mathrm{min}(\nu_0\Lambda\mathrm{CDM})-\chi^2_\mathrm{min}(\nu_Mw\mathrm{CDM}+N_\mathrm{fluid})=-21.08$, showing that the $\chi^2$ does improve by more than the additional number of free parameters.

These parameters deviate strongly from their $\Lambda$CDM values and are statistically compatible with the results from previous literature, introduced in Section~\ref{sec:prelim}.
We note that in this restricted analysis, the neutrino mass sum is higher than the $0.4~\eV$ found in previous studies~\cite{Beutler:2014yhv,Wyman:2013lza,Battye:2013xqa,McCarthy:2017csu}, but the results agree within the large error bars.
Note our results are not directly comparable to these previous works, which did not allow for both an evolving dark energy equation of state and a varying neutrino mass sum simultaneously, and some of them used an earlier \textit{Planck} SZ cluster measurement.
We also find that $\Delta N_\mathrm{eff}$ is consistent with zero, and $(c_\mathrm{eff}^2,c_\mathrm{vis}^2)$ are unconstrained, indicating that these datasets are not sensitive to this model extension.

\subsubsection{Full Datasets}

\begin{figure*}
\centering
\includegraphics[scale=0.4]{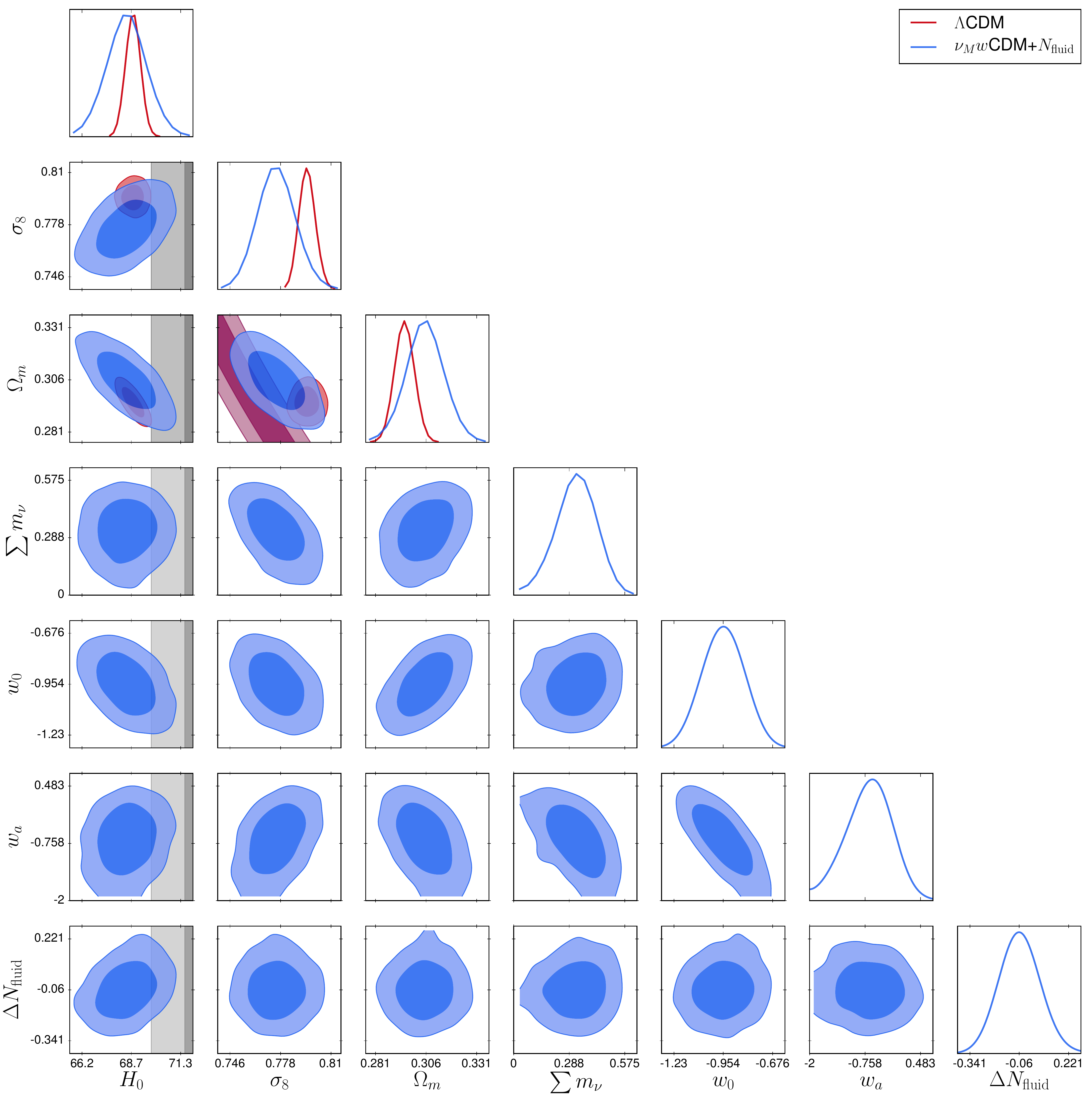}
\caption{The posterior distribution of  $\{H_0,\sigma_8,\Omega_m,\sum m_\nu, w_0,w_a,\Delta N_\textrm{fluid}\}$ when fitting to all datasets considered in this work, compared to the $\Lambda$CDM fit of the same dataset.}
\label{fig:LCDM_vs_mnuwCDM}
\end{figure*}

We turn to a full analysis that includes all datasets outlined in Section~\ref{sec:datasets}.
We compare the posterior distribution of $\{H_0,\sigma_8,\Omega_m,\sum m_\nu, w_0,w_a,\Delta N_\textrm{fluid}\}$ to that obtained in $\Lambda$CDM in Figure~\ref{fig:LCDM_vs_mnuwCDM}.
In Tables~\ref{tab:results_preliminary} and \ref{table:chi2_preliminary}, we report constraints on cosmological parameters, as well as the $\chi^2_\mathrm{min}$ contribution from each dataset.
These additional datasets restrict the ability of our ExDE model to resolve the tensions.
The BAO and JLA data, as shown in Table~\ref{tab:results_preliminary}, constrain the DE parameters to be very close to $\Lambda$CDM.
Additional ultra-relativistic species are still disfavored by the data: $(\Delta N_\mathrm{eff},c_\mathrm{eff}^2,c_\mathrm{vis}^2) = (-0.056_{-0.099}^{+0.093},0.53_{-0.3}^{+0.27},0.54_{-0.16}^{+0.46})$.

As a result, the central value of the $H_0$ measurement does not significantly change between the extended cosmology and $\Lambda$CDM.
The tension between the CMB and the SH0ES measurement is reduced to the $2.4\sigma$ level only because of the increase in error bars.
This is reflected in a modest change in $\Delta \chi^2_\mathrm{min}=-5.19$ with respect to $\Lambda$CDM at the expense of 5 new parameters.
The improvement to the fit is primarily due to a reduced $S_8$ tension between the CMB and the \textit{Planck} SZ data: $\Delta \chi^2_\mathrm{min} = -4.25$ from this dataset alone.
The parameter freedom that allows this improvement is the neutrino mass sum, which is measured as $\sum m_\nu = 0.32_{-0.09}^{+0.11}$.
Note that the $\chi^2_\mathrm{min}$ of the power spectrum measurements from SDSS and CFHTLenS is almost unchanged, indicating that they are consistent with this value of the neutrino mass.

In conclusion, it is possible to solve the $S_8$ tension with massive neutrinos even when the $H_0$ measurement is included in the analysis.
However, it is not possible to fully solve the $H_0$ tension within the $\nu_Mw$CDM+$N_\mathrm{fluid}$ model.
The values of $(w_0,w_a)$ required to make the SH0ES value of $H_0$ compatible with the CMB prediction are ruled out by BAO and supernovae, even when considering a combination of early- and late-Universe modifications.

\begin{table}[th]
  \begin{tabular}{|l|c|c|}
    \hline \hline
    Model &~~~~~~~~$\nu_0\Lambda$CDM~~~~~~~~& $\nu_Mw$CDM $+N_\mathrm{fluid}$ \\ \hline \hline
    $100~\omega_{b }$& $2.249_{-0.013}^{+0.013}$ & $2.229_{-0.016}^{+0.018}$ \\
    $\omega_\mathrm{cdm }$&   $0.1165_{-0.00076}^{+0.00075}$ & $0.1173_{-0.0018}^{+0.0017}$\\
    $100~\theta_s$    &  $1.042_{-0.00027}^{+0.00028}$ & $1.042_{-0.00087}^{+0.00066}$ \\
    $\ln 10^{10}A_{s}$  & $3.029_{-0.014}^{+0.011}$ & $3.042_{-0.019}^{+0.017}$\\
    $n_{s }$    & $0.9688_{-0.0038}^{+0.0036}$ & $0.9636_{-0.0053}^{+0.0055}$\\
    $\tau_\mathrm{reio }$  & $0.05133_{-0.0082}^{+0.0051}$ &$0.0578_{-0.0088}^{+0.008}$\\
    $\sum m_\nu$  & 0.06&  $0.32_{-0.09}^{+0.11}$\\
    $w_0$ & -1 &$-0.96_{-0.1}^{+0.11}$ \\
    $w_a$  & 0 &$-0.66_{-0.46}^{+0.52}$\\
    $\Delta N_\mathrm{eff}$  & 0 &$-0.0558_{-0.099}^{+0.093}$ \\
    $c_\mathrm{eff}^2$   & 1/3& $0.53_{-0.3}^{+0.27}$\\
    $c_\mathrm{vis}^2$  & 1/3&$0.54_{-0.16}^{+0.46}$ \\
    \hline
    $\sigma_8$  &       $0.795_{-0.0052}^{+0.0043}$ &$0.776_{-0.011}^{+0.011}$\\
    $\Omega_{m }$ & $0.2949_{-0.0044}^{+0.0042}$ & $0.3045_{-0.0088}^{+0.0087}$ \\
    $H_0$   &$68.82_{-0.36}^{+0.34}$ &$68.55_{-0.95}^{+0.96}$ \\
    \hline
  \end{tabular}
  \caption{Constraints at 68\% C.L. on cosmological parameters in various models including $\sum m_\nu$, $N_\mathrm{eff}$ and $(w_0,w_a)$ using all datasets considered in this work.}
  \label{tab:results_preliminary}
\end{table}

\begin{table}[th]
  \begin{tabular}{|l|c|c|}
    \hline\hline
    Model &~~~~~~~~$\nu_0\Lambda$CDM~~~~~~~~& ~$\nu_Mw$CDM $+N_\mathrm{fluid}$  \\ \hline \hline
    \textit{Planck} high-$\ell$ & 2460.67 & 2456.24 \\
    $\tau$ SIMlow & 0.24   & 0.17 \\
    \textit{Planck} lensing & 11.25 &11.32   \\
    SDSS DR7 & 45.77 &46.11 \\
    CFHTLenS & 97.92 &98.60 \\
    BAO (DES1) $z\sim 0.8$& 0.01 & 0.01 \\
    BAO $z\sim 0.10-0.15$& 2.82& 2.82  \\
    BAO $z\sim 0.4-0.6$& 7.14 & 7.82  \\
    BAO Ly-$\alpha$+QSOs&8.71 & 9.40 \\
    JLA &683.95 & 683.94  \\
    SH0ES& 5.29 & 6.63\\
    \textit{Planck} SZ & 9.14& 4.89 \\
    \hline
    Total $\chi^2_\mathrm{min}$   & 3332.89 &3327.70\\
    $\Delta \chi^2_\mathrm{min}$  & 0 & -5.19  \\
    \hline
  \end{tabular}
  \caption{The best $\chi^2$ per experiment for the standard $\nu_0\Lambda$CDM model and the $\nu_Mw$CDM $+N_\mathrm{fluid}$ .}
  \label{table:chi2_preliminary}
\end{table}

\section{Minimally parametric reconstruction of the Dark Energy dynamics}
\label{sec:agnostic}

In Section~\ref{sec:combine}, we restricted possible DE dynamics to those allowed by the simple $(w_0, w_a)$ parameterization of the DE equation of state.
We found that this parameterization did not allow enough freedom in the expansion rate to reconcile BAO and local $H_0$ measurements.
In this section, therefore, we consider what expansion rate would be required.
We use a fully general, minimally parametric model for the ExDE density as a function of redshift.
This allows the expansion rate to change essentially arbitrarily as a function of redshift.
In particular the expansion rate can match that expected for $H_0 = 69$ at $z > 0.15$, and thus match BAO, and then match $H_0 = 72$ at $z=0$.
We emphasise that the best fit parameters may not necessarily be realizable in a physical model.
In this section we are interested in determining what the data requires, partly to allow an assessment of the relative plausibility of explanations based on experimental systematics.

We write the Hubble expansion rate as
\begin{equation}
  H(z) = H_0\sqrt{\Omega_m(1+z)^3+\Omega_r(1+z)^4+\Omega_\mathrm{ExDE}(z)} \ ,
  \label{eq:Hubble}
\end{equation}
where $\Omega_\mathrm{ExDE}(z)$ corresponds to an unknown exotic DE species with an arbitrary density and equation of state.

Note that we do not restrict $\Omega_\mathrm{ExDE}(z)$ to be positive.
This allows us to include complicated dynamics resulting from, for example, a reduction in matter density from decaying dark matter or curvature.
This ExDE sector is implemented by modifying the expansion rate module in the Boltzmann code \texttt{CLASS}~\cite{Blas:2011rf}.
We neglect perturbations in the exotic fluid and change only the background expansion rate.
$\Omega_\mathrm{ExDE}(z)$ is given by a cubic spline interpolated between a series of values at different redshifts, called $z_\mathrm{knots}$.
We place a weak prior on the energy density of the exotic fluid at the knots to be $|\Omega_\mathrm{ExDE}(z_\mathrm{knot})|<4$.
We have checked that our results are insensitive to this choice.
Larger values are ruled out by the CMB.

To prevent our spline fitting the statistical noise of each dataset, we perform cross-validation (CV)~\cite{Sealfon:2005em}.
It is a standard technique in machine learning, based on the idea that a successful theory should be predictive.
When minimizing the likelihood function, we incorporate a roughness penalty based on the shape of the spline function $F_\textrm{ExDE}$
\begin{equation}
  F_\textrm{ExDE} = \int_{z_\textrm{min}}^{z_\textrm{max}}(\Omega_\textrm{ExDE}(z))''dz \ .
\end{equation}
In practice, we minimize the following quantity
\begin{equation}
  {\cal M} = -\ln~\mathcal{L} + \lambda F_\textrm{ExDE} \ ,
\end{equation}
where $\lambda$ is chosen according to the CV procedure.
We remove part of the data and perform a parameter fit for several values of $\lambda$ on the remaining datasets.
The best-fit parameters obtained from this limited dataset are then used to compute the $\chi^2$ associated with the removed part.
The value of $\lambda$ that minimizes the $\chi^2$ calculated on the set of data not included in the runs is $\lambda\sim 0.1$.
We investigate whether or not it is possible to solve the $H_0$ and $S_8$ discrepancies, accommodating all datasets in Section~\ref{sec:datasets}, and we investigate how changes in the background evolution influence the measurement of the neutrino mass sum.
All analyses include the CMB, LSS, SH0ES, and Ly-$\alpha$ BAO datasets.
We show results of fits including only a single $z<1$ dataset, either the galaxy BAO or JLA, and a fit including them both at the same time.

\begin{figure*}
  \centering
  \includegraphics[scale=0.35]{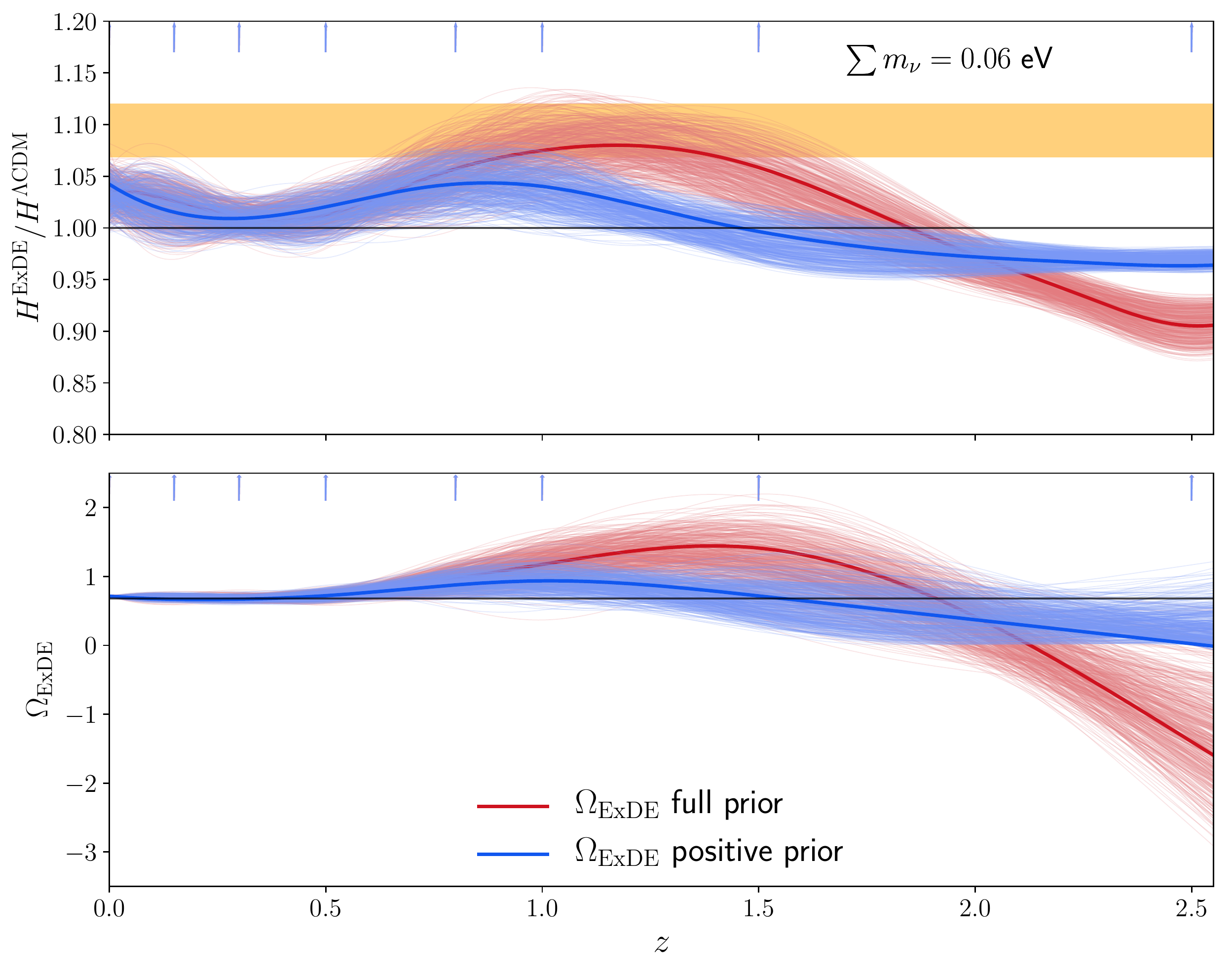}
  \includegraphics[scale=0.35]{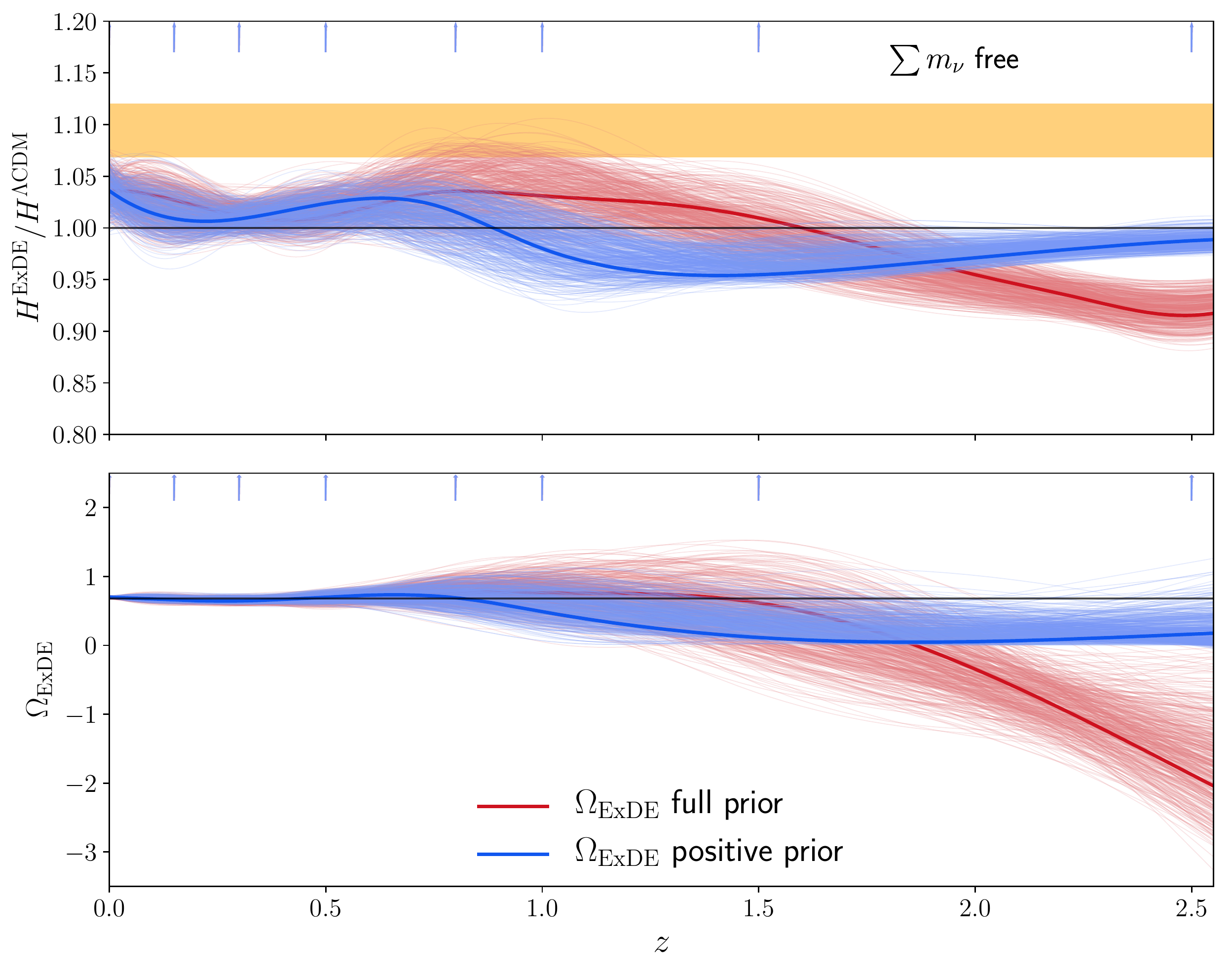}
  \caption{Reconstructed ExDE energy density and Hubble expansion rate (compared to the $\Lambda$CDM prediction from Planck TT,TE,EE+SIMlow, black line) with $\sum m_\nu=0.06~\eV$ (left panel) or $\sum m_\nu$ left as a free parameter (right panel), when including all datasets considered in this work and for different choice of prior on $\Omega_\mathrm{ExDE}$ (see text). 
    The thick solid lines show the best fit spline in each case, while the thin lines show samples from the $68\%$ confidence region. 
    The vertical arrows show the positions of the knots.  The orange band indicates the uncertainty on the Hubble parameter as measured by SH0ES (strictly speaking it is only valid a $z=0$). }
  \label{fig:reconstruction_AllData}
\end{figure*}

\subsection{Reconstruction from all datasets}

Since we use CMB data, we include a knot at $z = 1100$ and a knot at the initial redshift considered in \texttt{CLASS}, namely $z=10^{14}$, whose only purpose is to ensure a smooth interpolation.
We also include a knot at $z=0$ for the $H_0$ data and at $z=2.5$ for the Ly-$\alpha$ BAO.
The remaining knots are spaced linearly at low redshift and logarithmically at high redshift: $z=(0.15, 0.3, 0.5, 0.8, 1.0, 1.5)$.
Our knots are chosen based on the positions of each dataset, but our CV procedure dynamically reduces the number of degrees of freedom by correlating neighboring knots.
Thus, we expect that, as long as a sufficient number of knots are used, the positions and number of these knots will not affect our results once the CV roughness penalty is imposed. We discuss the robustness of our results in sec.~\ref{sec:robustness}.

Figure~\ref{fig:reconstruction_AllData} shows the best-fit curves for the late-Universe expansion rate $H^\mathrm{ExDE}$ (normalized to $\Lambda$CDM, using \textit{Planck} TT,TE,EE+SIMlow~\cite{Aghanim:2016yuo}) and reconstructed energy density $\Omega_\mathrm{ExDE}$ as a function of $z$, along with $500$ curves chosen at random from the $68\%$ confidence region.
The left panel shows the result with the neutrino mass sum set to $\sum m_\nu=0.06~\eV$, while the right panel shows the result with $\sum m_\nu$ as a free parameter.
We show expansion histories in which the neutrino mass sum is set to $\sum m_\nu = 0.06$ and those in which it is a free parameter.
We also show reconstructions which enforce a positive value for $\Omega_\mathrm{ExDE}(z)$ and those which allow $\Omega_\mathrm{ExDE}(z)$ to be negative.

$\Omega_\mathrm{ExDE}(z)$ is roughly constant when $\Omega_\mathrm{ExDE}(z) > 0$ is enforced.
However, when $\Omega_\mathrm{ExDE}(z)$ is allowed to be negative, the Ly-$\alpha$ BAO data make the best-fit $\Omega_\mathrm{ExDE}(z)$ negative for $2 \lesssim z \lesssim 2.5$.
The significance of this is greater than $68\%$, but does not quite reach $95\%$. This is unaffected by whether the neutrino mass is fixed, although fixing the neutrino mass causes an increase in energy density at $z  = 1.5$.
While it is possible that this could result from a modified gravity model, or potentially a decay in the dark matter density~\cite{Berezhiani:2015yta}, the most likely estimate is systematics in the Ly-$\alpha$ BAO data.
Note that by $z=1100$ $\Omega_\mathrm{ExDE}(z)$ is again positive, which argues against a cosmological explanation.
If we remove the Ly-$\alpha$ BAO, there is no data at $z=2.5$ and $\Omega_\mathrm{ExDE}(z)$ is consistent with zero and $\Lambda$CDM at this redshift.
Note that because the DR12 BAO likelihood is not yet public, we are using a Gaussianized version, which may underestimate the errors.
The best explanation for this discrepancy thus appears to be statistical.

\begin{figure*}
  \centering
  \includegraphics[scale=0.3]{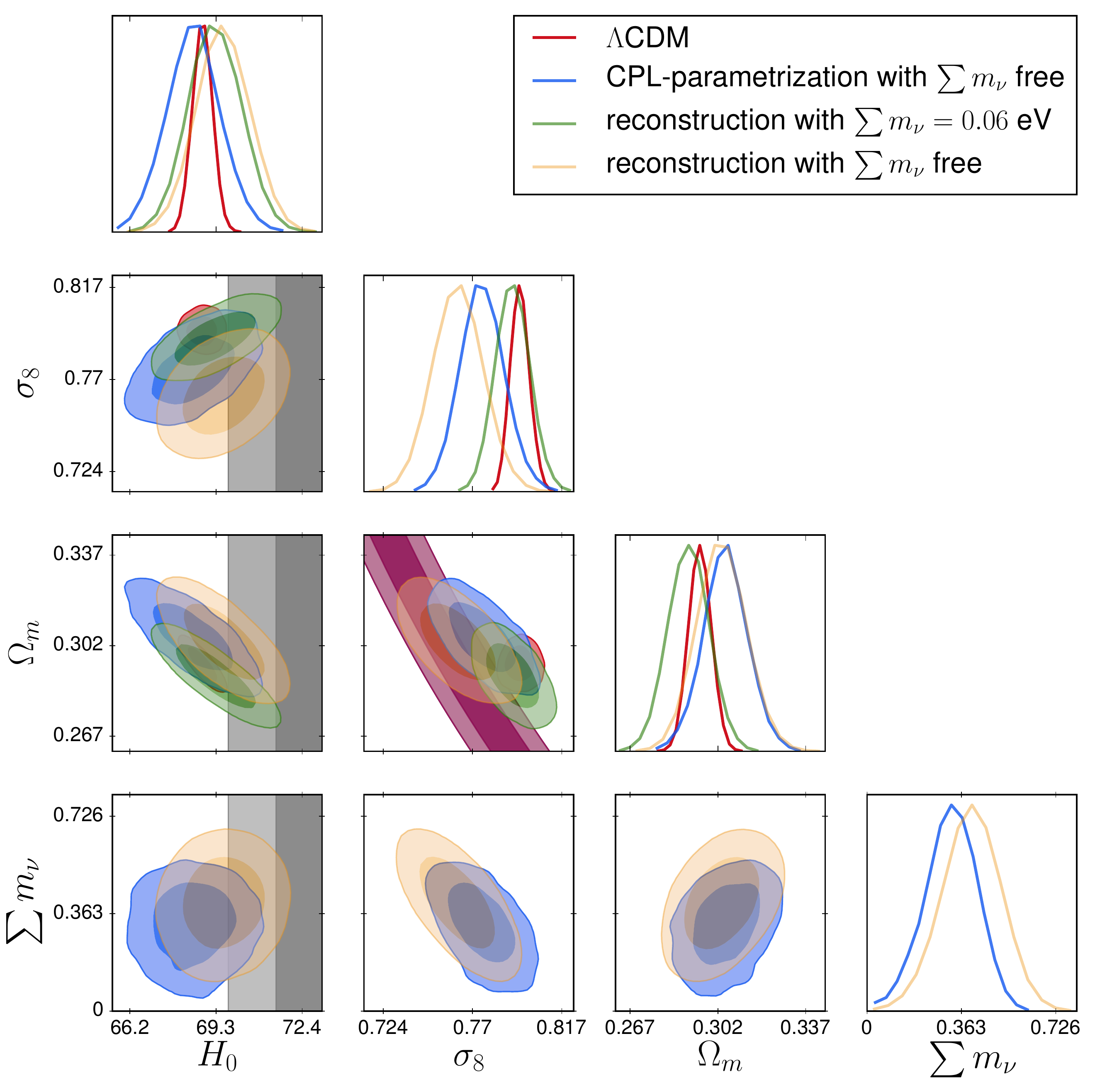}
  \includegraphics[scale=0.35]{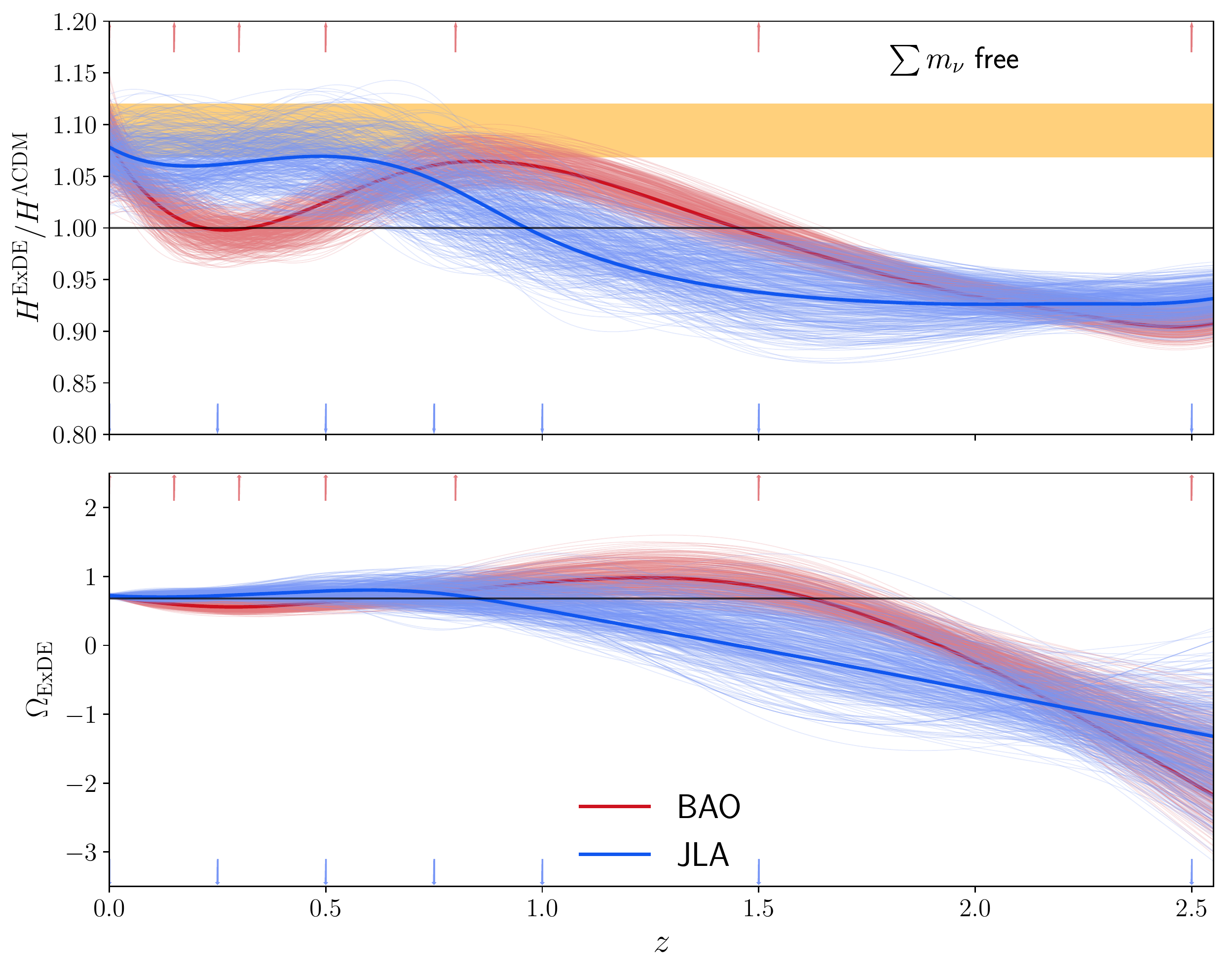}
  \caption{\textit{Left panel:} A comparison between the 1D and 2D posterior distributions of ($\sigma_8,\Omega_m,H_0,\sum m_\nu$) obtained in various models when using all datasets considered in this work.
    The grey band shows the R16 measurement, the purple band is the Planck SZ determination of $S_8$.
    \textit{Right panel:} Reconstructed DE energy density and Hubble expansion rate (compared to the $\Lambda$CDM prediction from Planck TT,TE,EE+SIMlow, black line) with $\sum m_\nu$ left as a free parameter.
    We include either the BAO (red) or JLA data (blue). 
    The thick solid lines show the best fit spline in each case, while the thin lines show draws from the $68\%$ most likely fits.
    The red arrows pointing upwards show the locations of the BAO knots, while the blue arrows pointing downwards show the positions of the JLA knots. The orange band indicates the uncertainty on the Hubble parameter as measured by SH0ES (strictly speaking it is only valid a $z=0$).}
  \label{fig:triangle_lcdm_vs_mnu}
\end{figure*}

If we weaken the effect of the Ly-$\alpha$ BAO data by, for example, enforcing $\Omega_\mathrm{ExDE}(z) > 0$, we see that the expansion history is consistent with $\Lambda$CDM within the error bars.
Thus, even when arbitrary DE dynamics are allowed, the tension between $H_0$ measured by SH0ES and that measured by BAO and the CMB remains.
Note however that the increased freedom in the model means that the tension is significantly weakened to less than $\sim1.9\sigma$.
One reason for this is that, given the value of $H_0$, the JLA and galaxy BAO measurements are in slight ($1-2\sigma$) tension.
This is illustrated in Figure~\ref{fig:triangle_lcdm_vs_mnu}: at $z\lesssim 0.6$ each experiment pulls $\Omega_\mathrm{ExDE}(z)$ in a slightly different direction, forcing an overall compromise value close to that of a cosmological constant.
The JLA data generally agree with the local $H_0$ data, while the BAO measurements agree with that from the CMB.
We emphasize that there is not necessarily any tension beyond statistical variation between these datasets.
Their agreement is well within the $2\sigma$ level.
The different behavior is mostly driven by the fact that fits to JLA data are insensitive to the value of $H_0$~\cite{Betoule:2014frx}.
Moreover, when combined together, their respective $\chi^2$ stays very good (see Table~\ref{table:bestfit_reconstructed_DE}).

\begin{figure*}
  \centering
  \includegraphics[scale=0.35]{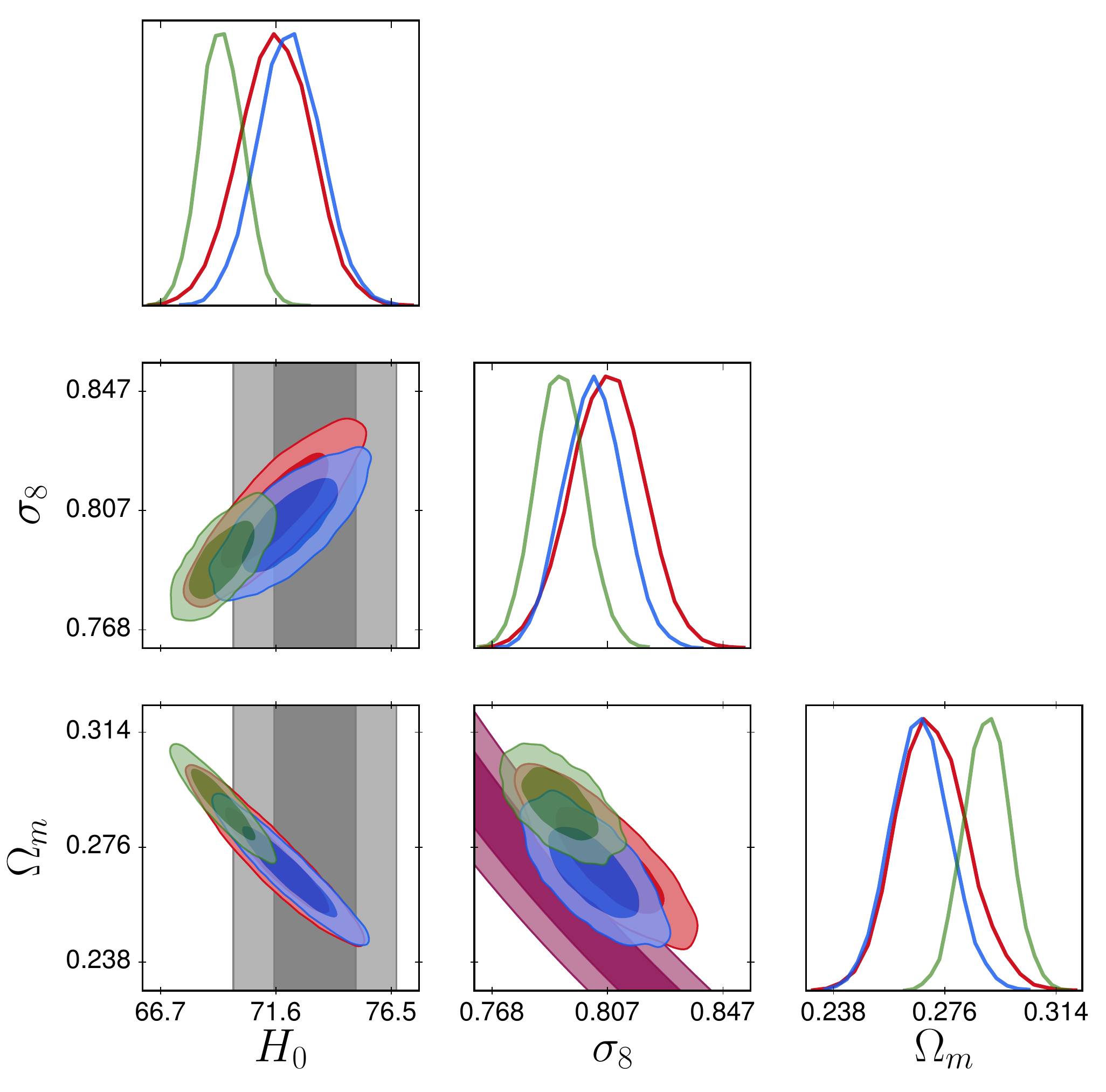}
  \includegraphics[scale=0.35]{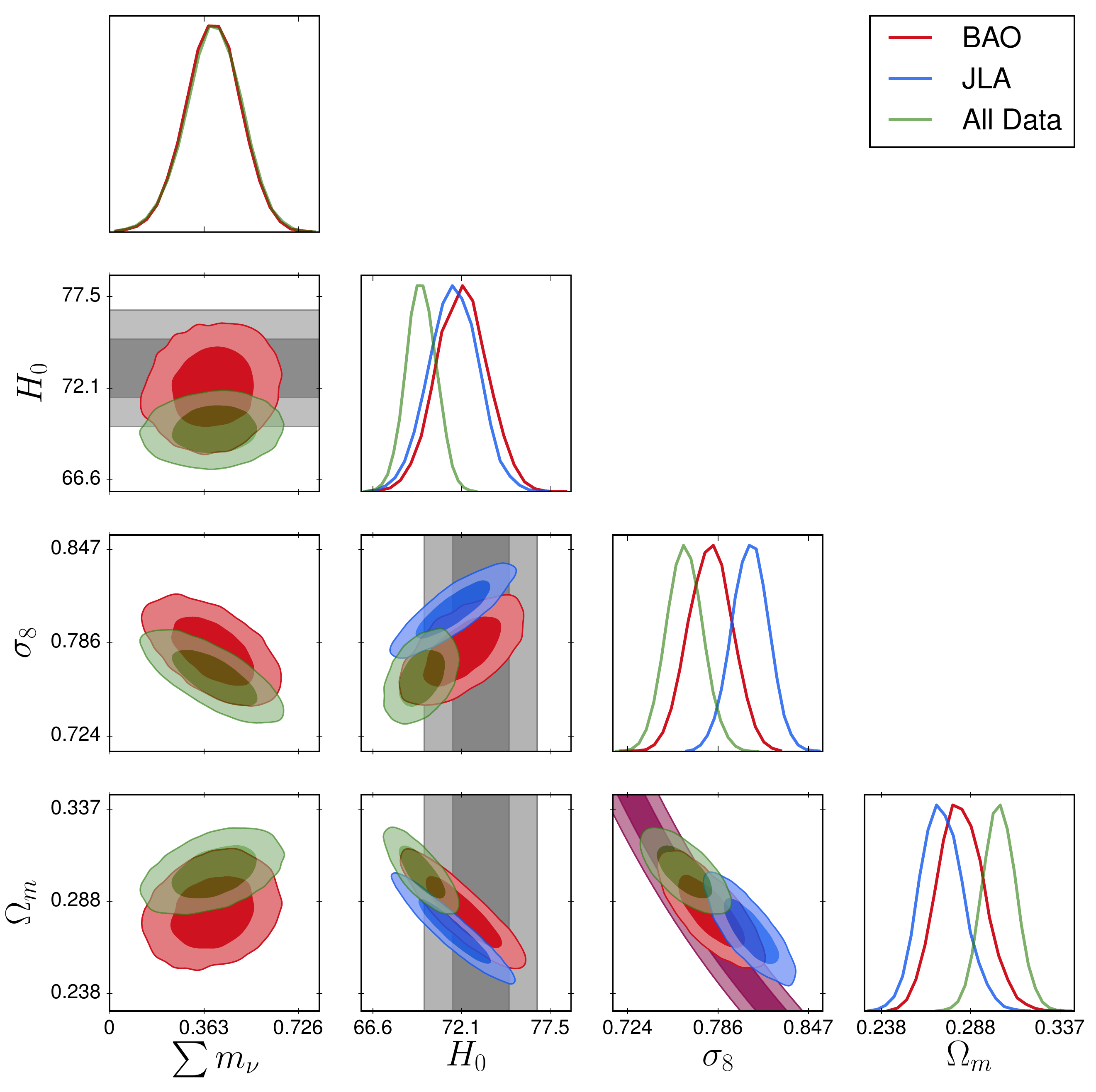}
  \caption{1D and 2D posterior distributions of ($\sigma_8,\Omega_m,H_0,\sum m_\nu$) with a fixed neutrino mass sum (left panel) and a free neutrino mass sum (right panel) when using SDSS DR7 CFHTLens, SH0ES, CMB, Ly-$\alpha$ BAO DR11, and either galaxy BAO DR12 (red curves) or JLA (blue curves).
    The grey band shows the SH0ES measurement, while the purple band is the \textit{Planck} SZ determination of $S_8$.}
  \label{fig:triangle_BAO_vs_JLA}
\end{figure*}

Interestingly, even with all datasets included, the neutrino mass sum is $\sum m_\nu=0.40_{-0.1}^{+0.11}~\eV$, driven by an improvement in the $\chi^2$ with the \textit{Planck} SZ data, as in Section~\ref{sec:combine}.
We have checked explicitly that the preferred neutrino mass changes by less than $1\sigma$ when omitting galaxy BAO or JLA from the datasets, even if the ExDE dynamics is very different from that of a cosmological constant.
This is illustrated in Figure~\ref{fig:triangle_BAO_vs_JLA}, where we show the posterior distribution of $\{\Omega_m,\sigma_8,H_0,\sum m_\nu\}$ obtained when allowing for a free neutrino mass (right panel) or fixing it to the minimal value indicated by oscillation experiments (left panel).

Table~\ref{table:bestfit_reconstructed_DE} shows the $\chi^2_\mathrm{min}$ for each dataset, fitting to $\Lambda$CDM, ExDE with the neutrino mass sum fixed to $\sum m_\nu = 0.06$, and ExDE with the neutrino mass sum left as a free parameter.
For each ExDE case, the prior $\Omega_\mathrm{ExDE}(z)$ is either restricted to be positive or is allowed to take on its full range of positive and negative values.
The Ly-$\alpha$ data near $z \sim 2.35$ are better fit with the ``full prior'', pulling $\Omega_\mathrm{ExDE}$ to negative values: the $\chi^2_\mathrm{min}$ in the ``full prior'' case is improved compared to the ``positive prior'' by $\Delta \chi^2_\mathrm{min}=-5.88$ when $\sum m_\nu=0.06~\eV$ and by $\Delta \chi^2_\mathrm{min}=-9.08$ when the neutrino mass sum is left free.
Finally, we perform an analysis of all datasets, including an extra ultra-relativistic fluid $(\Delta N_\mathrm{eff},c_\mathrm{eff}^2,c_\mathrm{vis}^2)$ and letting the neutrino mass sum vary.
We find that this additional ultra-relativistic species does not reduce the tension further, nor does it affect the reconstruction at low-$z$ or the determination of the neutrino mass sum.

\begin{center}
\begin{table}
\scalebox{0.8}{

  \begin{tabular}{|l|c|c|c|c|c|}
    \hline
    Model & ~~$\Lambda$CDM~~ & \multicolumn{2}{c|}{ExDE + $\sum m_\nu=0.06$}& \multicolumn{2}{c|}{ExDE + $\sum m_\nu$ free}     \\ \hline
    Prior on $\Omega_\mathrm{ExDE}$& $-$&~~~~~Full~~~~~&Positive&   ~~~~Full~~~~ & Positive  \\ \hline \hline
    \textit{Planck} lite  & 217.35 & 214.20&215.98 &  209.20&212.66\\
    $\tau$ SIMlow & 0.24& 0.06 & 0.06&  0.11&0.01\\
    \textit{Planck} lensing & 11.25 &10.03& 10.06 & 8.86&10.71  \\
    SH0ES& 4.75& 5.4& 3.32&4.28 &5.10\\
    \textit{Planck} SZ & 9.14 & 5.88 &8.64 & 0.12&2.58\\
    SDSS DR7 & 45.78 &44.97  & 45.05& 46.67&45.55\\
    CFHTLenS & 97.92& 97.06& 97.22&  97.90 &97.52\\
    DES1 BAO& 0.01& 0.05& 0.05& 0.01 &0.09\\
    BAO Ly-$\alpha$+QSOs &8.71& 3.88& 5.86 & 6.08&7.17\\
    BAO iso DR11& 2.81 &3.03 & 2.33&2.05 &2.39 \\
    BAO + $f\sigma_8$ DR12& 7.14& 4.08 & 4.11& 4.68 &5.37\\
    JLA &683.95&686.4 & 687.27 & 683.58 &684.85 \\
    \hline
    $\chi^2_\mathrm{min}$ & 1089.58&  1075.05&1079.93 &1064.70 & 1074.01 \\
    $\Delta \chi^2_\mathrm{min}$ & 0& -14.53 & -9.65&-24.88 & -15.57 \\
    \hline
  \end{tabular}}
  \caption{The best $\chi^2$ per experiment for the reconstructed DE dynamics with and without the neutrino mass sum as an extra free parameter when all datasets are included.}
  \label{table:bestfit_reconstructed_DE}
\end{table}
\end{center}

\subsection{Robustness of the result}
\label{sec:robustness}
We have performed a number of additional tests to assess the robustness of our conclusions.
First, we have checked explicitly that our results are robust to the addition of an extra high redshift knot at $z\sim4$.
As expected, we find that adding knots at this redshift and higher has no impact.
Indeed, there are no datasets sensitive to such redshifts (except for the CMB in a very mild way through the integrated Sachs-Wolfe effect).
Moreover, our prior on $\Omega_\textrm{ExDE}$ ensures that the Universe is largely matter dominated at these times.
We have also made several alterations to the position of the low-redshift knots [e.g. we set them at $z=(0.1,0.25,0.5,0.75,1,1.5,2.5)$] which had no significant effects on the reconstruction.
Additionally, we replaced the cubic spline with linear interpolation to check that our results are insensitive to our choice of parameterization.

Second, we have tested the robustness of our results to the addition or removal of datasets.
We find that our results are robust against exchanging the \textit{Planck} lite likelihood for the full likelihood.
Although we did not implement the full KiDS and DES likelihoods for this analysis, we checked that when the data from these experiments are reduced to a Gaussian prior on $S_8$ our best-fits are fully compatible with these measurements.
On the other hand, when removing the \textit{Planck} SZ likelihood, we find that $\sum m_\nu<0.48~\eV$ (at the 95\% confidence level) with a best-fit around $0.2~\eV$, indicating that $\sum m_\nu\sim0.4~\eV$ is perfectly allowed.
Moreover, following Ref.~\cite{Addison:2017fdm}, we have tested the possibility of removing \textit{Planck} data and using BBN data instead.
As expected, doing so has no strong impact on the late-Universe reconstruction; it simply increases the uncertainty on the densities of the various components in our Universe and reduces the $H_0$ tension to $\sim1.7\sigma$.

Finally, we have tested our results by introducing the extra free parameter, $A_\textrm{lens}$, which rescales the global amplitude of the lensing potential~\cite{Calabrese:2008rt}.
Ref.~\cite{McCarthy:2017csu} found that this can affect the constraining power of the lensing likelihood on $\sum m_\nu$.
We still find $\sum m_\nu = 0.31_{-0.11}^{+0.11}$, in very good agreement with our previous fit within error bars.
We additionally find $A_\textrm{lens} = 1.092_{-0.043}^{+0.041}$, in agreement with the value found by the \textit{Planck} analysis~\cite{Aghanim:2016sns}.
This value is discrepant at $2\sigma$ with the expected $\Lambda$CDM value of $1$ and thus represents an internal tension in the \textit{Planck} data due to an extra smoothing of the CMB high multipoles, as argued previously.

\section{Conclusions}
\label{sec:discussion}

In this paper we have examined two well-known tensions in the $\Lambda$CDM cosmology: the tension between local measurements of $H_0$ and the CMB-inferred value, and the tension between CMB measurements of the power spectrum amplitude $\sigma_8$ and that measured by galaxy clusters in \textit{Planck} SZ.
Many papers have focused on possible systematic explanations for these tensions.
We have instead assumed zero systematic error and investigated what models are required to explain these tensions taken at face value.
We show the 2D posterior distributions of $\{\Omega_m,\sigma_8,H_0,\sum m_\nu\}$ in the left panel of Figure~\ref{fig:triangle_lcdm_vs_mnu} for the various cosmological models considered in this work when including all datasets.

We first examined whether these tensions could be resolved by the simultaneous adoption of standard extensions to $\Lambda$CDM.
These extensions include massive neutrinos, extra relativistic degrees of freedom, and a fluid model of dark energy parameterized by a power law equation of state.
Several authors have previously used these extensions individually to resolve these tensions, but we consider enabling them at once.
We find that none of the extensions significantly reduce the tensions, with the exception of massive neutrinos.
We find that the addition of extra relativistic degrees of freedom does not reduce the tensions.
Since the galaxy BAO and JLA data measure the expansion history at relatively low redshift, there is insufficient freedom in the power law equation of state to reduce the tension with local $H_0$ measurements.

We found that a neutrino mass sum of $0.4~\eV$ could resolve the $S_8$ tension, and this resolution persists for the datasets we considered, as long as a model with enough freedom to reduce the significance of the $H_0$ tension was used.
The extra model freedom is important, because a side-effect of a non-zero neutrino mass sum is that it increases the tension between local $H_0$ measurements and the CMB by decreasing the inferred value of $H_0$ from the CMB.
However, a non-zero neutrino mass sum is well-motivated theoretically.
Whenever the $H_0$ tension is solved or greatly decreased, the $S_8$ value from \textit{Planck} SZ cluster count drives the neutrino mass sum to be close to $0.4~\eV$.
Remarkably, this result does not depend on the exact solution to the $H_0$ tension, which indicates that it is relatively robust.

Since explaining the total sum of cosmological datasets requires additional freedom in the expansion history, we included an exotic dark energy sector, which we allowed to have an energy density varying arbitrarily with redshift.
We emphasize that although we have assigned this sector to dark energy, it can be viewed as a proxy for other more physically motivated models, such as decaying dark matter or curvature.
We have not attempted to identify these models, treating the exotic dark energy sector as a purely phenomenological parameterization of the expansion rate.
We use cross-validation to avoid over-fitting the data.
We found that the best-fit model when all datasets was included was an expansion history relatively close to $\Lambda$CDM.
Thus the $H_0$ tension was not fully solved, although the extra model freedom did reduce the significance of the tension to less than $2\sigma$.
In order to fully solve this tension, it was necessary to also omit either the JLA data or the galaxy BAO data.
Either dataset allowed for a non-$\Lambda$CDM expansion history solution, but these solutions were inconsistent with each other.

We found that the Ly-$\alpha$ BAO dataset preferred a negative density of exotic dark energy at $z \sim 2.3$, a behaviour that cannot be recovered with an equation of state.
This result is not so cosmologically bizarre as it at first seems: for example, it could potentially be explained by an open Universe with a negative curvature component.
Although curvature is highly constrained by the CMB, these constraints are dependent on assuming $\Lambda$CDM and weaken significantly with more general models.
The presence of a negative curvature, as is the case if the Universe presents an open geometry, can naturally lead to apparent negative energy density for the dark sector.

Another possibility is that the exotic dark energy sector could include a decaying dark matter component. If the decay products dilute faster than matter, the expansion rate can be reduced around $z\sim2.3$.
However, the simplest such model, a dark matter component decaying into dark radiation with constant lifetime~\cite{Berezhiani:2015yta,Enqvist:2015ara}, is in conflict with observations of the late integrated Sachs-Wolfe effect and lensing power spectrum~\cite{Poulin:2016nat,Chudaykin:2016yfk}.
Moreover, we find $\Omega_\mathrm{ExDE}$ becomes positive again at $z < 1.5$.
Thus any decaying component must be accompanied by a later increase in energy density, tuned to restore agreement with $\Lambda$CDM.
Given that the negative energy density is driven by one dataset, some systematic in the measurement or moderate under-estimate in the error bars of the Ly-$\alpha$ BAO, is by far the most likely explanation.
To accommodate the data, $\Omega_\mathrm{DE}$ would then need to follow a dynamics very close to that obtained when restricting the analysis to positive priors on $\Omega_\mathrm{ExDE}$.
Such behavior can be obtained from a scalar field with a peculiar phantom behavior.
Of course, it would be theoretically more appealling to find a solution for which this behavior is not due to decoupled sectors, but arise from the common dynamics of several species related to each other.   Measurements of the expansion history at redshifts higher than those currently probed (for instance via future intensity mapping or 21cm BAO experiments) can allow us to understand whether the preference for exotic dark energy is real. If this behavior persists at higher redshifts, it can give important insights on the dark sector. However, if it does not continue, it can cast serious doubts regarding the validity of this interpretation of the Ly-$\alpha$ measurement.

While even our most general ExDE model was unable to solve the $H_0$ tension, there are classes of solutions not considered here.
For example, a modification of gravity such as Horndeski's theory~\cite{Horndeski:1974wa}, gravity theories with higher derivatives (e.g.~$f(R)$ gravity~\cite{Nunes:2016drj}, tele-parallel" $f(T)$ gravity~\cite{Nunes:2018xbm} or Galileon gravity \cite{Barreira:2014jha,Renk:2017rzu}) or nonlocal gravity (\cite{Belgacem:2017cqo}). The recently discussed ``redshift remapping'' is another potential solution that is not covered by our reconstruction \cite{Wojtak:2016dvd}.
Our reconstruction can serve as a guide to build a model, successfully explaining all datasets, and we may examine this in a future study.
Finally, we note that it is interesting that, whenever the $H_0$ tension was solved or weakened, the best fit neutrino mass sum was around $0.4~\eV$.
Future LSS surveys, such as Euclid and SKA, would be extremely sensitive to such a value of the neutrino mass sum~\cite{Sprenger:2018tdb}.

\section*{Acknowledgements}
We thank Joe Silk for interesting discussions.
This research project was conducted using computational resources at the Maryland Advanced Research Computing Center (MARCC).
Part of this work has been done thanks to the facilities offered by the Universit\'e Savoie Mont Blanc MUST computing center.
SB was supported by NASA through Einstein Postdoctoral Fellowship Award Number PF5-160133. This work was supported at Johns Hopkins by NSF Grant No.\ 0244990, NASA NNX17AK38G, and the Simons Foundation

\bibliography{biblio}

\end{document}